\documentclass[10pt,letterpaper]{article}
\usepackage[top=0.85in,left=2.75in,footskip=0.75in]{geometry}

\usepackage{amsmath,amssymb}

\usepackage{changepage}

\usepackage[utf8x]{inputenc}

\usepackage{textcomp,marvosym}

\usepackage{cite}

\usepackage{nameref}
\usepackage[colorlinks=true,linkcolor=blue,citecolor=blue,urlcolor=blue]{hyperref}

\usepackage[right]{lineno}

\usepackage{microtype}
\DisableLigatures[f]{encoding = *, family = * }

\usepackage[table]{xcolor}

\usepackage{array}

\newcolumntype{+}{!{\vrule width 2pt}}

\newlength\savedwidth

\raggedright
\usepackage[aboveskip=1pt,labelfont=bf,labelsep=period,justification=raggedright,singlelinecheck=off]{caption}

\makeatletter
\renewcommand{\@biblabel}[1]{\quad#1.}
\makeatother

\usepackage{lastpage,fancyhdr,graphicx}
\usepackage{epstopdf}
\fancyheadoffset[L]{2.25in}
\fancyfootoffset[L]{2.25in}
\newcommand{\ie}{{\em i.e., }}
\newcommand{\eg}{{\em e.g., }}
\newcommand{\etal}{{\em et al.}}

\begin{document}
\vspace*{0.2in}

\begin{flushleft}
{\Large
\textbf\newline{How the Avengers assemble: Ecological modelling of effective cast sizes for movies} 
} 
\newline
Matthew Roughan,
Lewis Mitchell,
Tobin South
\\
\bigskip
ARC Centre of Excellence for
    Mathematical \& Statistical Frontiers (ACEMS), \\ 
    School of Mathematical Sciences, University of Adelaide,
    Adelaide, 5005, SA, Australia
\medskip
\{matthew.roughan,lewis.mitchell,tobin.south\}@adelaide.edu.au

\end{flushleft}
\section*{Abstract}

The number of characters in a movie is an interesting
feature. However, it is non-trivial to measure directly. Naive metrics
such as the number of credited characters vary wildly. Here, we show
that a metric based on the notion of ``ecological diversity'' as
expressed through a Shannon-entropy based metric can characterise the
number of characters in a movie, and is useful in taxonomic
classification. We also show how the metric can be generalised using
Jensen-Shannon divergence to provide a measure of the similarity of
characters appearing in different movies, for instance of use in
recommender systems, \eg Netflix. We apply our measures to the Marvel
Cinematic Universe (MCU), and show what they teach us about this
highly successful franchise of movies.  In particular, these measures
provide a useful predictor of ``success'' for films in the MCU, as
well as a natural means to understand the relationships between the
stories in the overall film arc.



\section*{Introduction}

The Marvel Cinematic Universe (MCU) is arguably the most successful
movie franchise of all time. It has grossed more revenue from the
films alone\footnote{The MCU includes TV series series, one shot short
  movies and other media, but here we regard only the canonical series
  of 22 released movies (as of May 2019). See later for a precise
  list.} than the next two most successful (Harry Potter and Star
Wars).

What makes the MCU so successful? Clearly, there are many factors: the
deep and expansive nature of the source material they draw on, the
clever adaptation to movie format, the talented actors and directors,
and the number of pre-existing fans. However, there are other similar
attempts to translate graphic novels to the big screen, none as
large or successful as the MCU. 

The cast of a movie has a very important impact on its success. Most
obviously, ``star power'' can attract an audience, and ultimately the
talent of the actors and director are critical. However, underlying
this is a question of just how big the cast is. On a superficial
level, this seems trivial to answer, you just count. But that is
naive. Movie credits are not a uniformly defined source of data: the
onscreen credits of \emph{Closer} (2004) list 6 cast, whereas
\emph{Stuck on You} (2003) credits 383 actors. These numbers hide the
fact that the realistic count of meaningfully contributing characters
in these movies is similar.

Further, not all characters in a movie are equal. Most obviously, there
are named and unnamed characters. The first are those who have the
importance to require a proper noun designation, the second require a
designation but are not important enough to warrant a real name and
can include characters without lines such as those in the background
of a scene. Counting these so-called extras is problematic because
many of them are unlikely to be mentioned by name in the credits. And
more importantly, should a profusion of extras carry the same weight
in a metric of cast size as the number of named characters?
\emph{Ben-Hur} (1959) won a reputation as an epic in part because of
its reputed 10,000 extras, but should these each count as much as its
star Charlton Heston?

We argue in this paper that an \emph{effective} measure of the cast
size is useful, and such theoretically appealing metrics can be
derived from those used to measure ecological
diversity\footnote{The term used in
  ecology is diversity (or biodiversity), with reference to species,
  genera or similar, but the term diversity in analysis of media
  usually refers to gender or race, which is not the topic of this
  paper so we prefer to use alternative nomenclature.}~\cite{daly18:_ecolog_diver}. Such a metric
could be useful in
\begin{itemize}
    \item better predictors for movie success;
    \item taxonomic classification of movies for subsequent analysis; and 
    \item in creating features for recommender systems.
\end{itemize}

The metric we focus on here is a effectualisation of Shannon
entropy. However, entropy (and other metrics) are derived from a
probability distribution. The second major platform of this paper
is that the probability distribution underlying the metric is of great
interest: we investigate two here, one based on dialogue and the other
on the relative prevalence in \emph{conflicts} within the film.

We will show that the Shannon effective cast-size metric can measure
the size of a cast in a way such
\begin{itemize}
    \item that it have an intuitive meaning (\eg meaningful units);

    \item that it be built on firm mathematical foundations;

    \item that it be practically measurable, and that it be
      insensitive to noise in the data collection process; and

    \item that it is \emph{effective}, \ie numerically comparable across movies,
      and in particular useful for prediction or classification
      tasks. 
 
\end{itemize}

Thus we can think of the metric as an effective population size
metric, where population refers to the number of characters in a
movie. Moreover, this metric shows a useful relationship in
classifications of the movies both into generic types, and into
classes such as dialogue-based vs action-based. 

\section*{Related work}

Quantitative analysis of stories in most cases necessarily begins from a \emph{content analysis} of the text of the story itself.
Studies can be broadly categorised into two main approaches:
\emph{Temporal} studies tend to focus on aspects internal to the progression of the story, such as the `arc' of the narrative,
while \emph{structural} studies focus on the relationships between characters, often through social network analysis.
The present work fits most closely with the \emph{structural} approach as it focusses on measuring the effective number of characters,
however we borrow some techniques from information theory that are most typically associated with the \emph{temporal} approaches.

\emph{Temporal analysis of stories} --- While analysing plot structure has an extremely long history, arguably dating back to Aristotle (the Poetics), in modern (Hollywood) filmmaking the notion of the ``three act structure'' characterised by rising tension followed by resolution is codified/popularised by Syd Field \cite{field2005screenplay}. 
Taxonomies of stories more generally have been debated throughout the 20th century, with characteristic numbers of plots/stories found by authors apparently decreasing over time, from thirty-six in 1916 \cite{polti1921thirty}, to twenty in 1993 \cite{tobias1993twenty}, to seven by 2004 \cite{booker2004seven}.
Entropy has been extensively used in analysis of language, \eg see
\cite{brown92:_estim_upper_bound_entrop}.
Indeed, Shannon's original paper introducing information theory discusses redundancy in Joyce's ``Finnegan's Wake'',
and later non-parametric entropy estimators have been employed to quantify entropy in literature \cite{kontoyiannis1997complexity}, plays and poems \cite{rosso2009shakespeare}, and online social media \cite{bagrow2019information}.
Perhaps most similar to the methodological approach in this work are the studies on vocabulary \cite{d1991toward} and on placing information-theoretic bounds on human language acquisition \cite{mollica2019humans}, where entropy is used as a measure of language capacity.
Such approaches use information-theoretic techniques to estimate an ``effective'' vocabulary size,
in a similar manner to how we estimate the effective cast size here. 

At a more fine-grained level of temporal analysis over the course of a story,
Vonnegut proposed studying the ``shapes of stories'' by charting ``ill fortune-great fortune'' along a ``beginning-end'' axis \cite{vonnegut1999palm}.
This notion was made quantitative by analysing story ``arcs'' (which we distinguish from ``plots'' here)
by applying sentiment analysis techniques to the text of written works.
Notable examples here include the body of work by Jockers and coauthors \cite{jockers2013macroanalysis,JockersSyuzhet,archer2016bestseller,Gao2016},
as well as Reagan \etal, who find six characteristic emotional arcs in novels \cite{reagan2016emotional}.
For films, there is some suggestion that these emotional arcs relate to the eventual success of the film, as measured through box office returns \cite{del2018data}.
In this paper we find relationships between statistics of MCU films and box office returns as well,
but through our analysis of effective cast size.

\emph{Structural analysis of stories} --- The present work is more closely related to the literature on analysing structural properties of stories using networks, either between characters or portions of text.
Authors have studied the complex networks encoded in stories from throughout history, from ancient myths and sagas \cite{mac2012universal,Vikings}, to Romantic novels \cite{min2016narrative,ferraz2017representation}, to films \cite{ramakrishna2017linguistic}.
Of particular relevance here are recent social network analyses of the Marvel (comic) social network \cite{alberich2002marvel} and some social networks from superhero movies including \emph{Wonder Woman} and \emph{Thor} \cite{Jones2018}.
Edwards \etal \cite{edwards2018one} compare different ways of constructing social networks based on scripts, focussing on the television series \emph{Friends}.

To some extent this paper unifies the two major approaches to the quantitative analysis of stories by using entropy, which is typically associated with temporal analysis, to quantify something structural about the films we study, namely the effective cast size.

\section*{Materials and methods}

\subsection*{The anatomy of a cast list}

We are interested here in the cast of a movie, its \emph{dramatis
  person\ae} so to speak, but we are interested in the story, not its
means of implementation, so we are interested in the number of
characters (or parts or roles), not the number of actors (the two are
not identically the same). The most obvious place to access the list
of characters is through the credits.  A movie is constructed by a
long list of creative artists, actors and technical experts. Their
contribution to the movie is generally acknowledged through the
\emph{credits}, which historically have been displayed on screen at
the beginning of a movie in the title sequence and near the end in the
end credits. The credits are incredibly important to today's movie
participants because they effectively form their curriculum vitae.

However, there is no unique definition of how a credit list is
constructed, and it has varied historically and regionally in level of
detail, and the types of activity that are credited\footnote{For
  example, since \emph{Toy Story} (1995), lists of ``Production
  Babies'' born to crew members have been increasingly included in
  movie credits.}. In the United States, the standardisation of
credits starts around the 70s\footnote{\emph{Superman} (1978) had the
  longest credits at its release, setting a trend for superhero
  movies, whose credits can now include many thousands of
  participants.}, but has evolved continuously.

In the context of this paper the credits are important because one of
their key components is a cast list. However, credited cast lists
are often incomplete and sometimes misleading. There are many reason
for this, and discussion requires some understanding of the different
types of roles within a cast. Definitions for such vary (often base on
local union specifications) but losely we speak of
\begin{itemize}
\item Main roles: these are the primary roles in a movie. There are
  usually only a small number of such roles, and these characters
  perform the majority of action and dialogue.

\item Bit parts: these are smaller roles, often defined by the number
  of lines of dialogue the character has (less than 6 lines seems a
  common standard).

\item Extras: (or background talent, or atmosphere, or
  supernumeraries) these are much smaller roles, often as simple as
  providing background in a scene. The number of extras varies
  tremendously, potentially into the thousands. Extras are usually
  ``silent'' roles, though in the UK they are allowed to speak up to
  12 words).

\item Cameos: these are small roles, usually played by well-known
  actors or personalities (Stan Lee performed a cameo in all of the
  Marvel movies).

\item Doubles: stunt, body or costume doubles for a main actor fill
  the same character role with a different actor for various reasons.

\end{itemize}
The categorisation is important because credit lists of the cast of a
movie focus on the more important roles. Extras, cameos and doubles
may often be omitted for various
reasons~\cite{uncredited,uncredited2}.  Actors may not want to be
listed for artistic reasons (to enhance anonymity of a character where
that is important) or personal reasons, and movie producers may not
want spoilers to be disclosed through a cast list (these days cast
lists are often available before the movie release). Producers may
also not wish to credit actors because this may have implications for
employment costs, or simply because it dilutes the meaning of being
given credit. There are also many new types of participation modes for
actors (\eg voice only, or motion capture), and movies appear in
multiple formats (with different cutting) and it is ambiguous how
these might be credited, \ie should a character who only appears in a
deleted scene be credited?

IMDb collects cast lists for movies, somewhat provided by fans
identifying actors on screen, or actors self-identifying, and although
such contributed information can have inaccuracies, it allows a window
into how many roles are credited or
uncredited~\cite{uncredited,uncredited2}. IMDb also lists additional
attributes for roles: {\em uncredited, voice, motion capture, archive
  footage, scenes deleted, or credit
  only}~\cite{imdb_attributes}. Thus
the data is valuable, but it still has many issues. Most notably,
consistency is not enforced between movies, so a character (or even an
actor) might be listed under different names (see below for more
discussion).

Within a cast list, we notionally divide the characters into
subclasses based on these notions, and the extra notion of ``naming'',
\ie we break the characters into the classes:
\begin{itemize}
\item {\em Named:} These are characters that are assigned a proper noun
  designation, e.g., ``Iron Man''.

\item {\em Unnamed:} These are characters that are only indirectly names,
  \eg ``Tony Stark's secretary'', or given a job name or other
  designation, e.g., ``SHIELD agent''.

\item {\em Minor roles:} These are characters that only appear as
  ``uncredited'' in lists such as that of IMDb, or are not noted on
  any list. Even as such, they may play a role, \eg as the victim of
  another character's actions. 

\end{itemize}

In general we use IMDb to build such lists primarily as a comparison
point to demonstrate the difficulty of using such for assessing cast
size, and to illustrate the issues of data quality towards assessing
cast size. Our main data will be informed by the \emph{content} of
movies, explicitly to avoid such issues.

\subsection*{How \underline{not} to measure cast size}

In ecological terms a ``richness'' metric just counts numbers of
species. Here it would be a direct count of the characters in a
movie. In ecological settings this is seen to be
naive~\cite{daly18:_ecolog_diver}. In this section we demonstrate that
such counts are naive here also.

\begin{table}[!t]
  \begin{adjustwidth}{-1.25in}{0in} 
    \centering
    \begin{tabular}{r|lrrrr}
Title & Type & Credited & Named & Budget (\$M) & Box Office (\$M) \\
\hline 
Ant-Man & origin & 67 & 30 & 130 & 138.0 \\
Ant-Man and the Wasp & sequel & 61 & 27 & 162 & NaN \\
The Avengers & team-up & 53 & 23 & 220 & 623.3 \\
Avengers: Age of Ultron & team-up & 67 & 34 & 191 & 429.1 \\
Avengers: Infinity War & team up & 56 & 42 & 321 & 665.0 \\
Black Panther & origin & 66 & 22 & 200 & 501.1 \\
Captain America: Civil War & team-up & 103 & 25 & 250 & 408.1 \\
Captain America: The First Avenger & origin & 95 & 25 & 140 & 176.6 \\
Captain America: The Winter Soldier & sequel & 73 & 27 & 170 & 228.6 \\
Captain Marvel & origin & 51 & 30 & 152 & NaN \\
Doctor Strange & origin & 31 & 22 & 165 & 232.6 \\
Guardians of the Galaxy & team-up & 72 & 26 & 170 & 270.6 \\
Guardians of the Galaxy Vol. 2 & sequel & 48 & 34 & 200 & 389.8 \\
The Incredible Hulk & origin & 72 & 16 & 150 & 134.5 \\
Iron Man & origin & 63 & 26 & 140 & 318.3 \\
Iron Man 2 & sequel & 61 & 30 & 200 & 312.1 \\
Iron Man 3 & sequel & 110 & 39 & 200 & 409.0 \\
Spider-Man: Homecoming & origin & 64 & 37 & 175 & 334.2 \\
Thor & origin & 50 & 29 & 150 & 181.0 \\
Thor: The Dark World & sequel & 39 & 28 & 170 & 206.4 \\
Thor: Ragnarok & sequel & 43 & 18 & 180 & 315.0 \\
\end{tabular}

    \caption{Facts and figures for the MCU movies.}
    \label{tab:numbers}
  \end{adjustwidth}
\end{table}

The numbers of credited and named characters for each movie are given
in \autoref{tab:numbers}.  There are many notable defects that one
observes, for instance with regard to the number of credited
characters: 
\begin{itemize}
\item \emph{Iron Man 3} apparently has the largest cast. The movie,
  however, is a relatively straightforward linear plot with the lead
  character (Tony Stark/Iron Man) being dominant in terms of scene
  time and action.

\item On the other hand, \emph{Avengers: Infinity War}, which has a
  cast that includes almost every\footnote{A notable exception is
    Hawkeye / Clint Barton.} prior superhero from the franchise has a
  rather intermediate credited cast.

\item Similarly, \emph{Thor: The Dark World} also has a rather small
  list cast, given it takes place over three settings (Asgard and
  Earth and the eponymous Dark World) with three corresponding sets of
  cast members.

\end{itemize}

\noindent There are also issues with regard to the number of named characters: 
\begin{itemize}
\item \emph{Thor: Ragnarok} is listed as having the second smallest
  set of named characters. As with the other Thor movies this takes
  place over several settings, each with its own cast, and it involves
  a team-up between Thor (and his usual team) and the Hulk.

\item \emph{Spider-Man: Homecoming} is listed as having one of the
  largest sets of named characters. In this case, there are indeed
  quite a few such, but they play very little if any role
  in the plot.

\end{itemize}
However, a more important problem with this metric is that it does not
provide much separation between movies. The majority of movies all
lie in a narrow band in the middle. This lack of differentiation
means the metric is not terribly useful.

\subsection*{Measuring a character's participation}

Ideally, we would like to measure each actor's contribution to a
movie. However, this is difficult. Measures such as screen time are
difficult to measure, and in any case don't necessarily take account
of whether a character contributes meaningfully during this time. Thus
we need proxy measures of each actor's scale of contribution.

Many movies are driven by dialogue: what the characters say informs us
of the plot and the attitudes and emotions of the characters. Other
aspects are important, their actions, the setting, the background
music and so on, but the dialgue is the platform around which these
aspects revolve. Hence, one approach to measuring the participation of
a character in the movie is to measure the number of lines of dialogue
they speak. 

This is not the only metric, or even uniquely the best. Some
characters may be laconic, or overly verbose. And the importance of
any given line may vary.  At a deeper level, however, not all movies
are driven by dialogue.  A musical or opera is driven by the music
(the dialgue and plot are often only there to provide connections
between the songs). Pornographic movies are driven by sex, with plot
and dialogue again providing (usually extremely minimal)
connections. Here we are concerned primarily with the MCU, which falls
into the specific superhero genre, and more generally into the action
movie genre. And action movies are often driven by conflict.  Thus
measuring the number of conflicts in which a character participates
will give us another measure of their level of participation in the
movie.

\subsection*{Effective population size measurement}

Details often make a character, and so smaller cast of more carefully
constructed characters might be better; or perhaps volume carries
weight? We need a way to measure cast size in order to consider which
of these hypotheses is more true.

Our goal is to find an effective metric for the ``population size'' of
the cast of a movie. There are several goals in forming such a metric,
described in the introduction. Luckily the problem has been considered
extensively in the context of measuring ecology diversity of a habitat
(for a review see \cite{daly18:_ecolog_diver}), and there are several
parallels. In ecology there is a need to understand not just how many
members of each species are present in a habitat, but to capture what
this means in terms of diversity. The problem is complicated by the
noise in estimating species numbers, and the underlying variation in
those numbers. For instance, a naive count of the number of different
species can be misleading if most of
those species are near extinction.

However, there are several ecological metrics, and their purpose is
different, so rather than choosing one without care, we will consider
here the goals of such a metric in \emph{axiomatic} terms
\cite{e10030261}. That is, we shall describe (mathematically) the
properties that such a metric should have in our specific context, and
thereby derive a suitable metric.

First, such a metric should be based on the proportion of contribution of
each actor, not other features (such as their name). Hence, we can
write the effective number as a function of the proportional
contributions of each character to the story, \ie if we write the
effective number as $\hat{N}$ then it is a function
$\hat{N}(p_1, p_2, \ldots, p_m) =\hat{N}({\mathbf p})$ where the
proportional contribution of character $i$ is $p_i$ and there are $m$
characters to consider. Given this definition we have the axioms.
\begin{itemize}
    \item {\bf Non-negative:} $\hat{N}(\cdot) \geq 0$. It makes little
      sense to report a negative population size. 

    \item {\bf Continuity:} $\hat{N}(\cdot)$ is a continuous function
      of the variables: small changes in the data should result in
      small changes in the metric. 

    \item {\bf Symmetry:} Reordering the actors, for instance such
      that $p_i$ and $p_j$ are swapped, should not change the
      metric. For instance
      \[ \hat{N}(p_1, p_2) = \hat{N}(p_2, p_1). \]
      This condition arises because we want a metric in which
      characters are interchangeable, in order to be able to make
      comparisons, for instance, between different movies. 
 
    \item {\bf Normalisation:} If there are $M$ equally contributing
      characters, \ie $p_1 = p_2 = \cdots = p_m = 1/M$, then we
      require that $\hat{N}=M$, and this should be maximal. This
      condition is imposed to scale the values intuitively, so that a
      value of $\hat{N}$ can be associated to an actual number of
      characters.

    \item {\bf Zeros:} Adding a character that makes no contribution
      should not change the metric, \eg
      \[ \hat{N}(p_1, p_2, 0) = \hat{N}(p_1, p_2). \]
      A net result of this and the prior condition is that
      $\hat{N}(1,0) = 1$. This condition arises both for technical
      reasons, and because we don't want noise (small errors) to lead
      to instability in the metric. For instance, introducing a
      character who makes very little contribution should not change
      the metric by much.

    \item {\bf Monotonicity:} If the number of 
      participating characters increases, then the metric should
      increase, \eg assuming all participation is non-zero
      \[ \hat{N}(p_1, p_2) < \hat{N}(p_1, p_2', p_3'), \]
      where $p_2' + p_3' = p_2$ where $p_2', p_3'>0$. The intention is
      that larger casts should report a larger metric.

\end{itemize}
Additionally we expect the metric to be sub-additive in the sense that
if one considers two movies jointly, then the effective cast size of
the joint movie should be no larger than the sum of the casts of the
two components.

This list of axioms has redundancies, but more importantly it does not
lead to a unique metric. There are several alternatives that are
valid. We use the method based on Shannon Entropy $H({\mathbf p})$ \cite{e10030261},
which is defined as follows:
\begin{eqnarray}
    H({\mathbf p}) & = & - \sum_i p_i \log_2 p_i, \\
    \hat{N}({\mathbf p}) & = & 2^{H({\mathbf p})},
\end{eqnarray}
with the common convention that $0 \log 0 = 0$. The metric
$\hat{N}({\mathbf p})$ has sometimes been called the \emph{perplexity}
of a distribution (because of its relationship to one's ability to
predict the outcome of a random event with this distribution) and has
been used in natural language processing
\cite{brown92:_estim_upper_bound_entrop}.

There are other metrics that satisfy these axioms\footnote{These
  roughly correspond to axioms used to derive Shannon entropy, but
  there is a key grouping/partitioning/recursivity axiom missing here
  as it appears hard to justify in the context.} \cite{e10030261}, but
the metric we have chosen has several advantages: (i) Shannon entropy
is easy to calculate, (ii) it is used in many other fields (\eg
physics and information sciences) and so has common interpretations,
(iii) the maximum entropy principle naturally leads to models, and
(iv) entropy has many generalisations, for instance to distance
metrics -- this is a key advantage here in that we can adapt the same
ideas for comparisons between movies.

The metric leaves open the means by which we estimate the $p_i$, the
proportional contributions of each character. We test the use
of dialogue or conflict.

\section*{Data}

\subsection*{Public data sources}

Public databases, e.g., IMDb and OMDb, are used here to provide cast
lists, revenue and cost figures, and ratings. Some considerable effort
has gone into cleaning the cast lists in particular. The most notable
problem is aliasing of character names, \ie a single character may
appear under different names in different movies. This problem is
particularly prevalent in superhero movies. Alias lists were construct
by hand for all significant characters in the movies.

As a source for dialogue, publicly available movie scripts were used. 
This has been a valuable data source in previous work
\cite{webb2010corpus}, but has been focused in constructing dialogues
for use in computational linguistics \cite{serban2015survey}. Here we
are not analysing the text of the speech, but rather the speakers. 

No single source has scripts for all of the MCU movies. We used two
sources: 
\begin{itemize}
\item The community run Transcripts Wiki on the website
  \emph{fandom.com} \cite{transcriptswiki}. These webpages were
  downloaded and processed to find appropriate lines of dialogue and
  its speaker. The scripts come in a number of formats, each requiring
  different methods of parsing. Some scripts had no consistent format
  or incomplete scripts (see below for more discussion).

\item PDF documents of movie scripts were sourced from
  Script~Slug~\cite{scriptslug} if they were unavailable or incomplete
  on the Transcripts Wiki.  The text from the PDFs was extracted and
  parsed to collect the names of each speaker of dialogue. Individual
  care was put into each PDF parsing as formats varied.
\end{itemize}

Transcripts, especially fan transcripts, often lack the depth of
information desired for more complex analysis
\cite{baldry2016multisemiotic}, but the extraction of speakers in
dialogue sequences is relatively reliable. However, as noted,
incompleteness was a problem.

Even with two data sources we did not have complete scripts for all
movies. Only 14 movies have complete scripts. Three have mostly
complete scripts that may miss a scene or have minor uncorrected
issues throughout the script. The four remaining movies had only
partial scripts. Details of the movies and numbers of lines of
dialogue available in their transcription are shown in
\autoref{tab:integrity}.

\begin{table}[th]
 \begin{adjustwidth}{-2.25in}{0in} 
    \centering
    \begin{tabular}{r|rrrl}
Title & Run Time (m) & No. of conflicts & No. of lines & Status \\
\hline 
Ant-Man & 117 & 82 & 865 & complete \\
Ant-Man and the Wasp & 118 & 110 & 180 & incomplete \\
The Avengers & 143 & 237 & 830 & complete \\
Avengers: Age of Ultron & 141 & 302 & 975 & complete \\
Avengers: Infinity War & 149 & 278 & 991 & complete \\
Black Panther & 134 & 117 & 728 & complete \\
Captain America: Civil War & 147 & 263 & 982 & complete \\
Captain America: The First Avenger & 124 & 99 & 619 & complete \\
Captain America: The Winter Soldier & 136 & 149 & 822 & complete \\
Captain Marvel & 123 & 165 & 686 & partial \\
Doctor Strange & 115 & 112 & 159 & incomplete \\
Guardians of the Galaxy & 121 & 157 & 576 & partial \\
Guardians of the Galaxy Vol. 2 & 136 & 127 & 956 & complete \\
The Incredible Hulk & 112 & 92 & 63 & incomplete \\
Iron Man & 126 & 81 & 124 & incomplete \\
Iron Man 2 & 124 & 112 & 1006 & complete \\
Iron Man 3 & 130 & 121 & 571 & partial \\
Spider-Man: Homecoming & 133 & 58 & 1558 & complete \\
Thor & 115 & 95 & 873 & complete \\
Thor: The Dark World & 112 & 106 & 732 & complete \\
Thor: Ragnarok & 130 & 197 & 970 & complete \\
\end{tabular}

    \caption{Size and completeness of datasets, including the numbers
      of transcribed conflicts and lines of dialogue, and whether the
      dialogue file was complete, partially incomplete or largely
      incomplete.}
    \label{tab:integrity}
  \end{adjustwidth}
\end{table}

\subsection*{Conflict transcription}

Underlying this work is the desire to form a theory of conflict in
narrative analysis. Much of narrative concerns conflict, for instance,
between a hero and villain. The development of this component of a
narrative is particularly obvious in the action movie genre, in which
plots of movies are predominantly carried by a sequence of
interdependent conflicts. Indeed, there are famous cases of movies
whose main actor spoke almost unintelligibly, but carried the film
through charisma and physical presence, which indicates the importance
of ``action'' to the genre. 
 
Mathematically, we can think of a story as having a set of characters
${\cal C}$. A conflict is a mapping from a pair of these characters to
an outcome, \ie it is a function $f_i(a,b) \rightarrow \{a, b, d\}$,
where $a,b, \in {\cal C}$ and $d$ means a draw or an indeterminate
outcome.  A scene is made up of a sequence of such conflicts
$i=1,\ldots,n$. For instance, a movie scene often follows a
backwards-forwards motif oscillating between minor victories for the
villain and hero. Alternatively, the hero might be the underdog,
physically over-matched by the villain, suffering repeated setbacks,
but eventually, the hero wins by adopting a clever stratagem. We are
not concerned in \emph{this} paper with the temporal structure of this
process. Here we are only interested in the proportionate
contributions of each character to the action of the story, but future
work aims to tease out such motifs from this data.
  
It is not always obvious from observation how to divide a story into a
sequence of such mappings. Technical considerations aid in the
decision: (i) if conflicts are pairwise, then we must divide conflicts
involving more than two parties into components small enough to be
modelled as pairwise, and (ii) estimates based on the data will be
limited by the resolution of the data. Both of these considerations
push towards as fine a division of conflicts as possible. On the other
hand, there is a natural desire that the outcome $d$ be in the
minority, \ie that most conflicts have a winner and loser, and this
means that we cannot meaningfully break a scene into individual
frames. Additionally, the overhead of a frame based approach is
impractical for manual transcription, not to mention that individual
frames are often meaningless without being seen in context. 

We transcribed the conflict data manually by viewing and annotating
the movies. Transcription was performed by pausing the movie
(frequently) so that notations could be made immediately, avoiding
problems of memory.  In transcribing we are aiming to balance the
alternative pressures: to subdivide scenes at a fine granularity vs
the desire for conflicts to have determinate outcomes.  There is
inherent subjectivity in any such transcription process. How finely to
subdivide conflicts, and even who is victorious is often
straight-forward, but not always. Ideally we would involve multiple
transcribers, and using a metric of inter-coder reliability determine
a consistent set of data. However, (i) the movies represent a large
corpus of material; (ii) conflict data does not fall into a standard
model for inter-coder concordance in any case\footnote{For instance,
  commonly used correlation coefficients are designed to assess
  agreements amongst ratings, not between choice of subdivision of a
  dataset.}, and (iii) we acknowledge that the theory of conflicts
described so far is not fully formed at present. Instead, we aimed
here to at least provide a dataset that was consistently transcribed,
and to build a metric of effective cast size that was robust to
transcription variations. 

However, in order to reduce inconsistency, all movies were examined
and transcribed by one observer. In addition, the order of
transcription was randomised (constrained by movie releases) to avoid
systematic biases.  Transcription was performed over as short a period
as practical for the same reason\footnote{The main body of the MCU was
  transcribed over a few weeks, but transcription of \emph{Captain
    Marvel} occurred somewhat later when it was released outside
  theatres. \emph{Avengers: End Game} has not yet been transcribed as
  it only has a theatrical release at present.}, again with the aim of
applying the same standard to classification.

A more difficult issue to address is that as transcription progressed,
there was a parallel learning process about how to most accurately and
consistently transcribe events. In particular, the level of
subdivision, and the degree of inclusion of conflicts between minor
parties increased in later transcriptions as their importance was
realised. In order to address this drift in transcription we
reprocessed the first three movies. The second transcription of these
movies was substantially different from the first, which contains a
much courser breakdown, and hence many fewer conflicts. Fortunately,
this data provides a comparison that allows us to test the sensitivity
of our metric to the collection methodology (see below).

In general, releases of movies in different regions can be cut
slightly differently, for instance to fit into a particular ratings
scheme.  The movies transcribed were all the standard Australian
release. We did not analyse ``director's cuts'' or other such modified
releases.  We have not seen any evidence that the MCU movies released
in Australia differed from the US release, or any other worldwide
releases.

\subsection*{Data cleaning}

\begin{quote}
  It’s often said that 80\% of the effort in a data analysis is spent
  on data cleaning, the process of getting the data ready to analyse.
\end{quote}
\vspace{-3mm}\hspace{23mm}  Hadley Wickham \cite{wickhamed:_tidy}
 
The conflict data described above was collected as expeditiously as
possible. It was quickly typed into a spreadsheet, resulting in
typographic errors. Likewise fan transcripts contained some
errors. These errors were detected by parsing the data, and validating
against a set of criteria, \eg time sequence monotonicity, name
frequency, and simple matching algorithms; as well as matching to
external data sources. Errors were corrected manually, with reference
to the original movie where the mistake was not obvious (typos were
generally very obvious once discovered).

A more critical issue lies in character names. Many characters appear
in multiple MCU movies. However superhero names are not standardised.
Characters introduced in one film under a given name are sometimes
reintroduced under another (\eg Bucky Barnes v the Winter Soldier).
Also naming conventions in movie credits are not standardised, and
superhero movies add a layer of complexity because superheroes often
have aliases. For instance Natasha Romanoff is known as the superhero
Black Widow, but also appears in credits as ``Agent Romanoff'' and
``Natalie Rushman'', and abbreviations or combinations of
these. Characters also sometimes appear in alternative forms, \eg
``Young Gamora,'' or are known by nicknames.  The dialogue in scripts
also uses variant forms of these names. An aliases file was created in
order to disambiguate characters and convert all appearances to a
common format. Extensive use was made of ``canonical'' sources of data
on the movies, in particular \cite{mcu_wiki} in order to ensure
accuracy of the alias lists\footnote{The listing is available at
  \url{https://github.com/mroughan/AlephZeroHeroesData/tree/master/MarvelCinematicUniverse}.}.
Nearly 200 characters are listed in this data, commonly with two
aliases, but sometimes with significantly more. For example, The Winter
Soldier, James Buchanan `Bucky' Barnes has six.

Unnamed characters do not use the alias list, and some generic
characters may be grouped together by a credits lists or the names
used in scripts. For instance, two separate characters may be listed
as ``SHIELD Agent''. The number of such, and the number of conflicts
and/or lines of dialogue that such characters appear in is small, but
may result in a marginal decrease in the effective cast size. A key
advantage of the approach proposed here is a high degree of robustness
to such noise.

All data and code will be released with the publication.

\subsection*{Types of movies}

In order to better understand results, the movies are classified into
types within the MCU based on the their characteristics and
relationship to other movies in the franchise. We classify movies as
\begin{itemize}
\item {\em origin} movies -- these are the first major appearance of a
  character, and often explain some of that characters back story;

\item {\em sequels} -- these are movies that follow on directly from
  another with a large overlap in cast and plot elements; and

\item {\em team-ups} -- these movies involve a group of superheroes
  forming into team for a larger purpose.

\end{itemize}
The classification is soft in the sense that many of the movies have
aspects of more than one type of movie: for instance, most movies in
the MCU are sequels in some sense. Other movies such as
\emph{Guardians of the Galaxy} is both a team-up and an origin movie
for the characters in the team. Here we identify the primary role of
the movie. The classification is given in \autoref{tab:numbers}.

\section*{Results}

\subsection*{Effective (conflict) cast size}

\begin{figure}[th]
    \centering 
    \includegraphics[width=\linewidth]{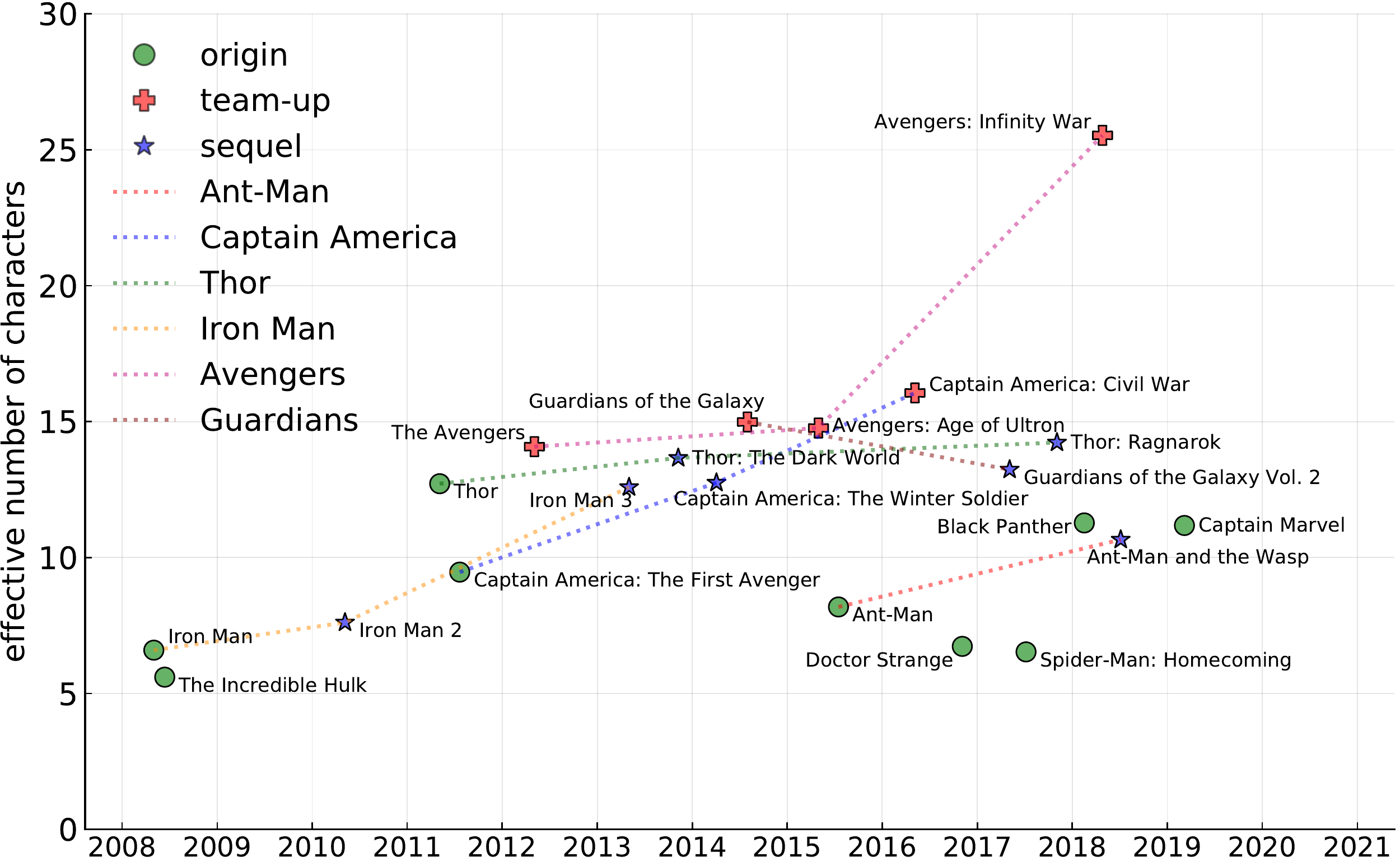}
    \vspace{2mm}
    \caption{Effective cast size of each movie in the MCU showing type
      of movies by shape, and sub-sequences connected by dashed
      lines. The $x$-axis is the theatrical release date.}
    \label{fig:effective}
\end{figure}

The movies considered are listed in \autoref{tab:numbers}.  We
consider first the effective cast sizes obtained from the conflict
data because of the completeness of this data.  Effective cast sizes
are shown in \autoref{fig:effective} with the type of movie indicated
by the marker, and specific sub-sequences of the overall set of movies
are indicated by dashed lines.  There are many interesting features of
this plot.

When considered by class we see notable features: most origin movies
have a small effective cast, which grows in sequels. The exceptions to
this rule are the Thor and Guardians sequences. 

The Thor movies have a more stable cast size (though the actual cast,
and even the distribution varies). The movie \emph{Thor} is unusual as
an origin movie because it takes place across two major settings
(Earth and Asgard) each with their own somewhat disjoint casts. It is
therefore not surprising that the effective cast size is approximately
double that of other origin movies.  The Guardians of the Galaxy
movies start with a team-up movie without the typical origin movies
for team members, making these movies an outlier within the MCU, but
leading to consistent casts sizes when considered by type. 

Another notable feature is a general level of \emph{inflation} in the
cast sizes with time. This is particularly the result of an increasing
number of team-up movies which have larger casts, and \emph{Infinity
  War}'s huge cast (this movie is a team-up of teams).

None of these results are terribly surprising. Most fans or movie
critics might have drawn a plot with a similar shape given a little
thought, but deriving this from actual cast lists is hard as we showed
earlier. It is a measure of the success of the metric that it should
match with intuition about the movies. 

We can also consider the overall cast size of the MCU. If we combine the
movies by taking averaged proportional contributions from the full
set of characters, then the effective cast size is approximately 119.6
characters. This is an extraordinarily large number for a franchise,
and reflects both Marvel's willingness to develop multiple streams of
movies building up to the central \emph{Avengers} movies, and the fact
that the action is well-distributed amongst this cast, \ie there is no
single, central character that dominates the movies.  Compare, for
instance, to the James Bond franchise which has a single, extremely
dominant character, the eponymous James Bond.

On that topic, the top-5 characters that make the largest contributions to
the overall metrics are in  order:
\begin{enumerate}
\item Iron Man / Tony Stark;

\item Captain America / Steve Rogers;

\item Thor Odinson;

\item The Hulk / Bruce Banner; and 

\item Black Widow / Natasha Romanoff.

\end{enumerate}
Again, there is no surprise here, but the relatively equal
contributions from the top-5 characters is somewhat unique in film
franchises. 

The large cast size of 119.6 characters is also interesting in
comparison to a simple sum over the effective cast sizes of the movies
(248.5). In ecological terms this disparity would be described in
terms of $\alpha$-diversity (the within habitat biodiversity,
$\beta$-diversity (the between habitat biodiversity) and
$\gamma$-diversity (the overall biodiversity of a set of habitats)
\cite{whittaker1972evolution}. The MCU sits in an interesting place
where a significant part of its diversity occurs inside each movie
($\alpha$), and a significant part between movies ($\beta$) but there
is also a strong overlap (which we will consider further later in the
paper).

\subsection*{Prediction}

\begin{quote}
I win my awards at the box office.
\end{quote}
\vspace{-3mm}\hspace{23mm}  Cecil B. DeMille

A metric is as useful as its uses. In the context of movie production
a key performance indicator is how profitable the movie is, \ie the
ratio of the movie's box office receipts\footnote{There are many
  other facets to the success of a movie: DVD sales, for instance, and
  Star Wars is legendarily famous for its merchandising, but Box
  Office receipts is (i) a somewhat easier number to obtain, and (ii)
  is possible to obtain more quickly than other metrics, which is
  important in order to compare recent movies.} to its
cost\footnote{Exact costs are notoriously hard to obtain -- here we
  use the budget as stated in IMDb.}. Some effort has gone into
developing predictive models for movie performance, \eg see
\cite{lash16:_early_predic_movie_succes}. Amongst many factors they
consider (audience-based, release-based, and content-based) the
content includes who is in the cast (that is the actor, not the
character), and the actors' popularity. They also note that features
of the acting team can contribute, but again they are considering
actors not cast, and they consider semantic features of the group (\eg
team chemistry) rather than purely quantitative measures.  Instead, we
test the effective cast size as a predictor: this is much simpler to
measure, as it is a feature of the movie itself, not the various
qualities and relationships of the actors.

\begin{figure}[th]
  \centering
  \includegraphics[width=\linewidth]{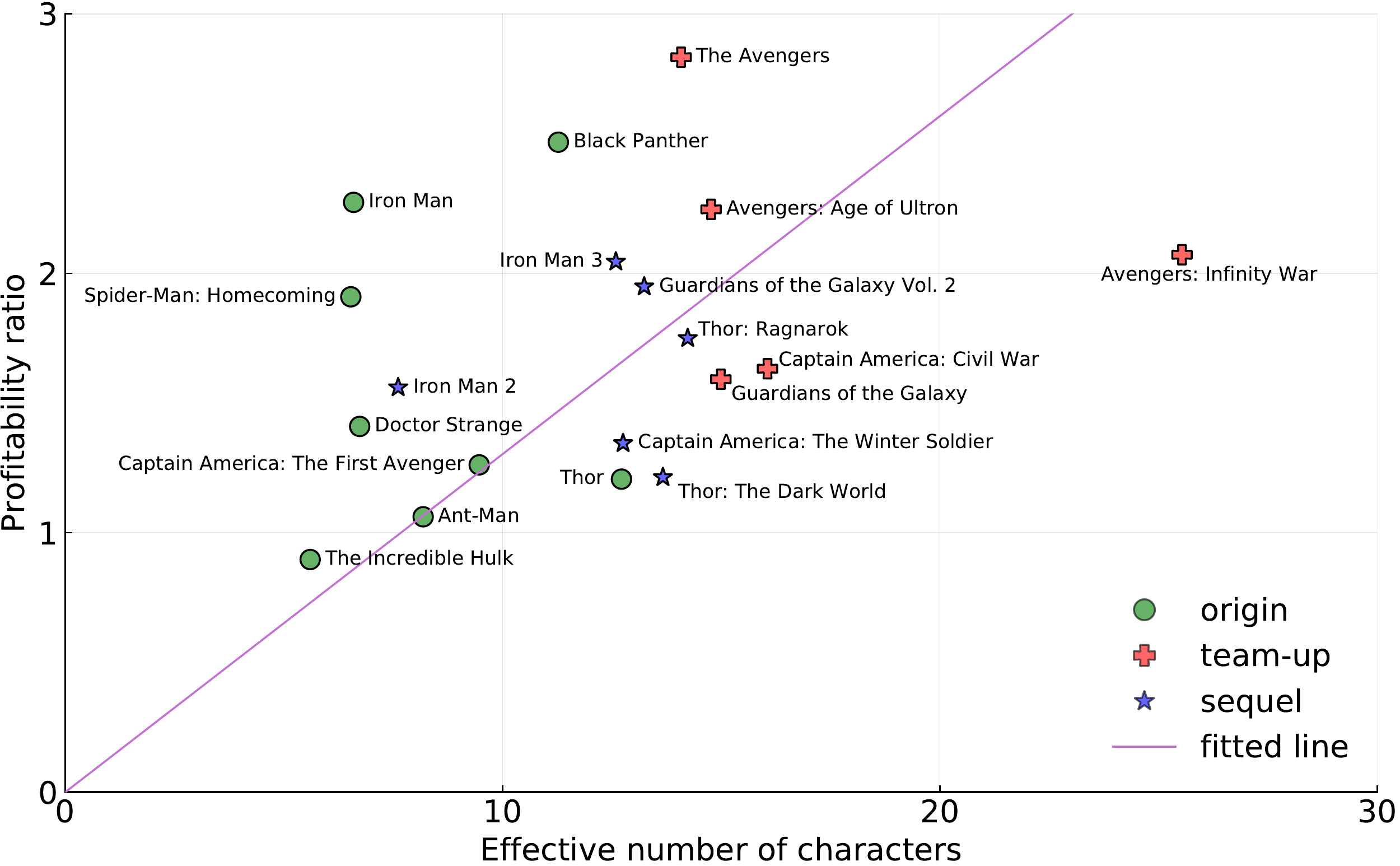}
  \vspace{2mm}
  \caption{Profitability as a function of effective cast size. }
  \label{fig:profit}
\end{figure}

In \autoref{fig:profit} we compare the effective cast number to the
profitability ratio. Of note in the MCU is that all but one movie have
ratio greater than 1, the exception being \emph{The Incredible Hulk},
which was widely considered a failure (its lead actors was one of the
very replaced in later movies). However, its role in establishing one
of the key characters of the franchise should not be underestimated.

We also plot on \autoref{fig:profit} a least-squares fitted
line\footnote{The fit is is through 0, the argument being that a
  movies with no cast, would have profitability zero.}. The $p$-value
for this fit is below $10^{-8}$ indicating strong significance to the
relationship, and Pearson's correlation coefficient is 0.184
indicating a non-negligible correlation between the two statistics.

There is considerable variation around the line. There are quite a few
movies that lie well above it: for instance \emph{Iron Man}, which has
been hailed as one of the best movies in the franchise, and as one of
the key initiators of the cinematic universe. Quality of acting and direction,
cast ``star power'', timing and other factors cannot be discounted as
important to the overall success of a movie. However, the effective
cast size also appears to influence the profitability of a movie. 

This is a fact that does not seem to be missed by the producers of the
MCU franchise. As we noted earlier, larger casts seem to be becoming
more common. 

Why should this be? Do audiences really prefer a larger cast? We have
to be very careful about this conclusion because the movies with
larger casts tend to be team-ups. Dedicated fans might watch all of
the MCU movies, but other fans may choose to watch only a single
sub-sequence. When these culminate in a team-up, this draws in
audiences from the multiple strands.

\emph{Guardians of the Galaxy} is a useful data point here. This movie
is ostensibly a team-up, but lacks the usual prequel origin
movies. Although the movie was widely praised, it wasn't a profit
generating engine compared to several other team-up movies.

The fact that the MCU is constructed not of individual movies, or
individual sequences is one of its chief beauties. There appears to be
an understanding in the producers of the franchise that the smaller
scale origin movies are a necessary pre-condition for the profit
generating team-up movies.

\begin{figure}[th]
  \centering
  \includegraphics[width=\linewidth]{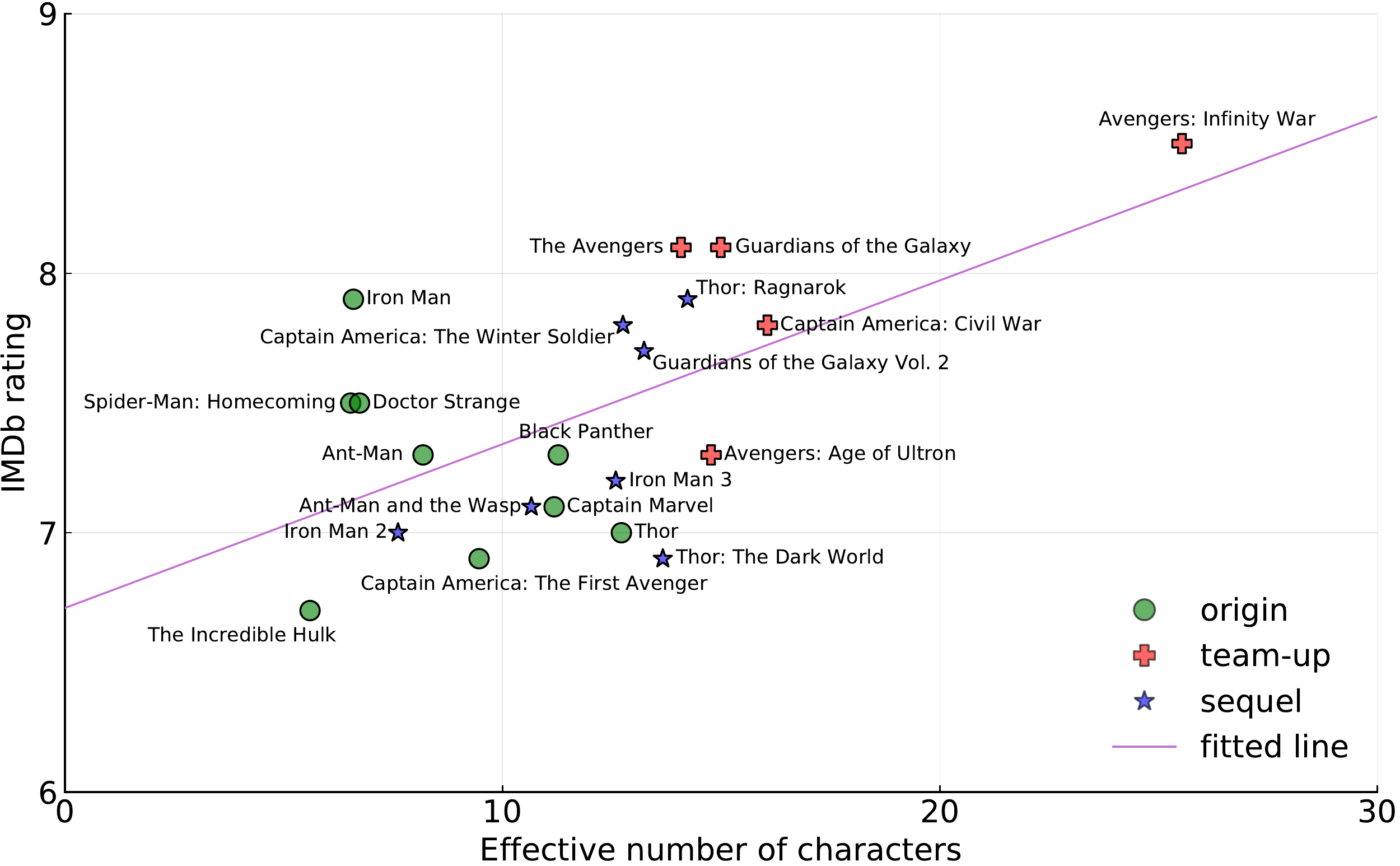}
  \vspace{2mm}
  \caption{IMDb rating as a function of effective cast size. }
  \label{fig:rating}
\end{figure}

We can also look at the relationship between ratings and cast size
(see \autoref{fig:rating}) and we see the same type of pattern. Larger
effective cast sizes are correlated with higher ratings, though once
again we must take care to note that this is a correlation, not
necessarily a causative relation.

\subsection*{Robustness}

A key concern in any such metric is the impact that noise in the underlying data might have on the results, particularly where there is an element of subjective judgement in the transcription process. 
To understand how sensitive the effective cast size is to transcription variations, we performed two separate transcriptions of three of the MCU movies. The two transcriptions were very different: the second set were performed at a much courser level. \autoref{tab:sens} shows the results. The number of conflicts found in each dataset is listed in order to show just how different these two transcriptions were. The effective cast sizes for each case are also reported. The notable feature of effective cast size is that it varies by a much smaller amount than the input data. There are some changes, mainly because in the courser grained analysis, some characters do not appear at all, and hence it makes sense that the effective cast size would decrease, but this is a comparitively small decrease! 

\begin{table}[th]
    \centering
    \begin{tabular}{r|rr|rr}
 & \multicolumn{2}{c}{Dataset 1} & \multicolumn{2}{c}{Dataset 2} \\
Title & No. of Conflicts & $\hat{N}$ & No. of Conflicts & $\hat{N}$ \\
\hline 
Iron Man & 81 & 6.59 & 36 & 4.74 \\
The Avengers & 237 & 14.08 & 92 & 13.41 \\
Avengers: Age of Ultron & 302 & 14.77 & 180 & 12.95 \\
\end{tabular}

    \caption{Table showing results for three cases with two
      alternative views}
    \label{tab:sens}
\end{table}
 
This robustness to input data variations is a key ingredient in any good metric.

\subsection*{Effective (dialogue) cast size}

So far we have only considered the conflict-based metric for cast
size.  In this section we consider the related dialogue metric -- as a
reminder this uses the same formulation but an alternative estimate of
the proportional contribution of each character to the story.  

\begin{figure}[th]
    \centering 
    \includegraphics[width=\linewidth]{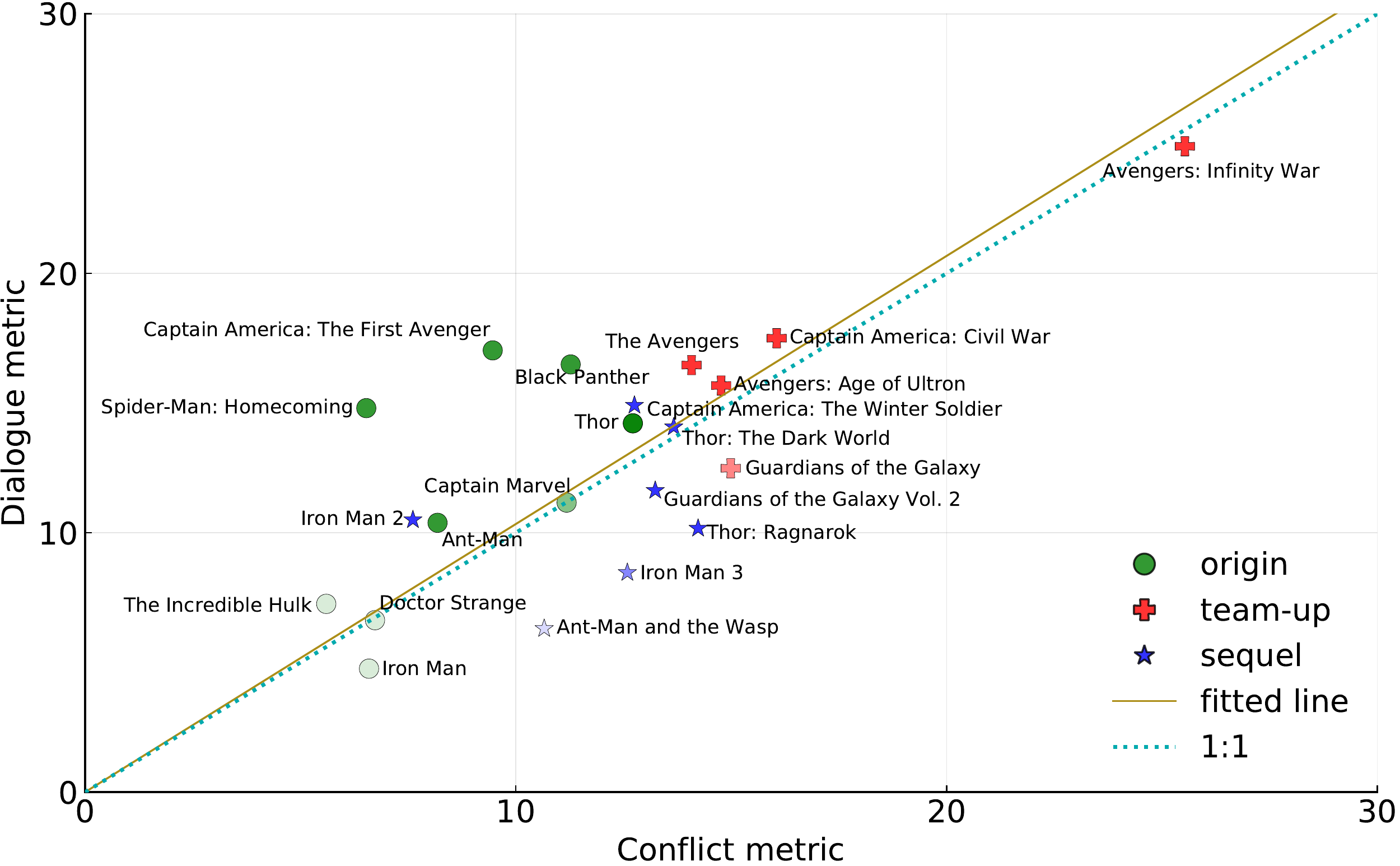}
    \vspace{2mm}
    \caption{Conflict and dialogue metrics of cast size. Shading
      indicates less complete (dialogue) datasets.}
    \label{fig:dialogue}
\end{figure}

\autoref{fig:dialogue} shows a scatter plot of the two alternative
metrics. The plot also shows two reference lines: the first a fit
(through zero) to the data, and the second a 1:1 line. The former
shows that overall the dialogue measures of cast size report very
slightly larger numbers than the conflict based metric. However, they
are remarkably close given the fact the two use completely different
views of the movie.

There are additional patterns of note. We can understand that movies
that sit above the reference line contain more dialogue-based
participation, and less conflict, and in turn, those below the line
entail more conflict. Extreme examples are \emph{Spider-Man} and
\emph{Captain America: The First Avenger}.

Origin movies usually lie above the line (or close to it). Origin
movies often use extensive dialogue to show character development from
``zero to hero.'' 

Sequels seem, on the other hand, to lie close to or below the line,
thus involving more action than dialogue.  Team-ups often lie quite
close to the line: the formation of a team often requires the members
to meet and talk, but these movies use action to display the abilities
of the entire team. 

The location of a movie in this plot represents subtle changes in
style in screenplay and direction. It is interesting that a level
of detail that one might expect only to be exposed by detailed
semantic analysis of content can be seen in a single pair of metrics.

\section*{Movie comparison results}
\label{sec:comparison}

An additional task for which cast metrics are useful is comparison of
movies. That is, we would like to have a metric for how similar or
dis-similar two movies are. For the same reasons that we consider
above, it would be useful to have a metric that is measured in units
that correspond to effective cast size, \ie it would be appealing to
say that the difference between two movies is $X$ effective cast
members. Thus we should have that two disjoint movies $A$ and $B$
would have an distance that is directly related to
$\hat{N}(A) + \hat{N}(B)$, and that two movies with identical
character contributions would have a distance of 0.

There are many formal distance metrics that can be applied to
probability distributions. A survey of metrics and their properties
can be found in \cite{gibbs02:_choos_metric}.  Typical distance
metrics between probability distributions either
\begin{enumerate}
\item assume the same support, \ie, that the two have positive support
  on an identical set of elements (\eg Kullback-Leibler or $\chi^2$
  divergences), or

\item they ignore the measures on the elements and simply use counts
  of the size or relative size of overlaps of support (\eg Jaccard,
  Bray–Curtis or Dice distances/dissimilarities).
 
\end{enumerate}
Neither of these two conditions is applicable in our domain.  We need
a metric that takes into account the probabilities, but also allows
the supports to be different as different movies have intrinsically
different casts. 

There are two approaches we can adopt: one formal, and the other more
intuitive. The formal approach is to adapt the Jensen–Shannon
divergence (JSD)
\cite{dagan97:_simil_based_method_for_word_sense_disam,lin91:_diver_shann}
(also called the \emph{total divergence to the average} or the
\emph{increment of the Shannon entropy}). This is a symmetrised,
smoothed version of the Kullback–Leibler divergence (KLD). It can be
defined in terms of the KLD
\[
  D_{KL}( P |\!| Q ) = - \sum_{x \in X} P(x) \log \frac{Q(x)}{P(x)},
\]
as 
\[ 
   D_{JS}( P |\!| Q) = \frac{1}{2} \left[ D_{KL}( P || M ) + D_{KL}( Q || M ) \right],
\] 	
where $M = (P+Q)/2$ is short-hand for the distribution formed from the
averages of the probabilities of the two distributions or equivalently
in terms of entropy
\[ 
   D_{JS}( P |\!| Q) = H\left( M \right) - \frac{1}{2}\big[ H(P) + H(Q) \big].
\] 	
It has the advantage of being theoretically understood \cite{dagan97:_simil_based_method_for_word_sense_disam,lin91:_diver_shann}, \eg we know
that $0 \leq D_{JS}( P |\!| Q)  \leq 1$. In linguistics
\cite{dagan97:_simil_based_method_for_word_sense_disam} this has been
used by taking an exponent, the natural such for us being to take a
power of 2, \ie 
\[ \bar{D}_{JS} = 2^{D_{JS}( P |\!| Q)} - 1. \]
We subtract 1 because we should like our measure of
dissimilarity to be 0 when comparing a movie to itself. We can form a
measure of similarity then by noting that the maximum value of
$D_{JS}$ is 1, and taking
\[ \bar{S}_{JS} = 1 -  \bar{D}_{JS}. \]

The approach above is theoretically appealing, consistent with our use
of entropic measures, and useful, but it lacks an interpretation in
terms of effective cast size. That is, we can only interpret $\bar{S}_{JS}$
as describing the \emph{proportion} of cast members in common.

An alternative measure is to proceed with the calculation we might
pursue in terms of measuring the effective increase in cast size of
the combination of two movies over the average size of the casts by
taking
\begin{eqnarray*}
   D_{\rm effective}( P |\!| Q)
    & = &   \hat{N}(M) - \mbox{average}\big[\hat{N}(P)  + \hat{N}(Q) \big] \\
    & \simeq &   2^{H(M)} - 2^{ (H(P) + H(Q)) / 2  }.
\end{eqnarray*}
This expression looks very similar to that for $\bar{D}_{JS}$ but 
it has units of  characters.

The two appear similar, and in fact are very closely related if we
normalise $D_{\rm effective}$, by taking
\[ \bar{D}_{\rm effective}( P |\!| Q) = \frac{ D_{\rm effective}( P |\!| Q)}{(\hat{N}(P) + \hat{N}(Q))/2}
\]
and as before we form a similarity measure $\bar{S}_{\rm effective} = 1 - \bar{D}_{\rm effective}$.

\begin{figure}[th]
    \centering 
    \includegraphics[width=0.8\linewidth]{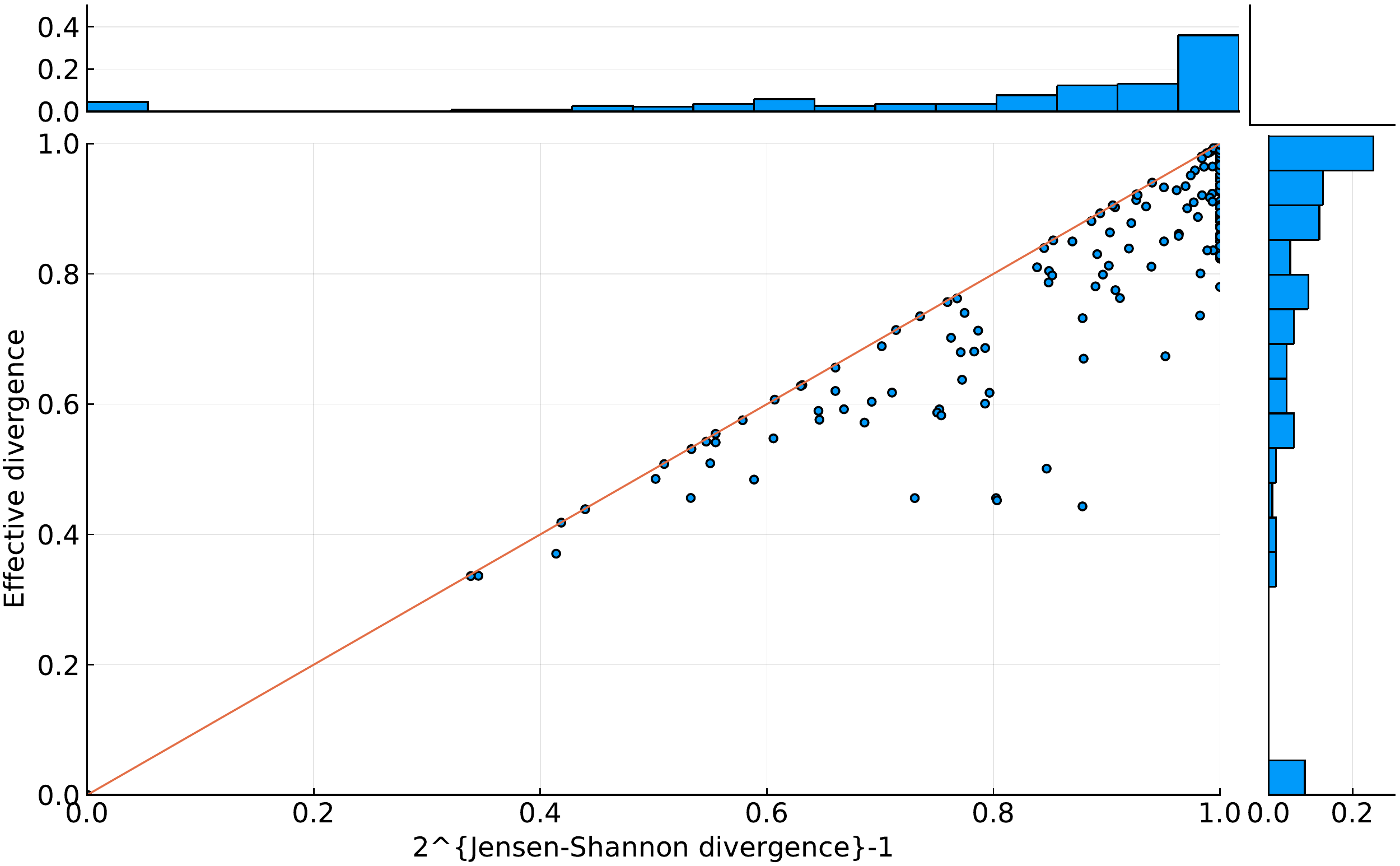}
    \vspace{2mm}
    \caption{A comparison of the two distance metrics showing that
      effective divergence $bar{D}_{\rm effective}$ lies below the
      Jensen-Shannon measure $\bar{D}_{JS}$.}
    \label{fig:comp_distance}
\end{figure}

The two have very similar results, but the latter is can be smaller
(as one might expect from Jensen's
inequality). \autoref{fig:comp_distance} shows a comparison between
the two. The effective measure is highly correlated with the
Jensen-Shannon measure, but has the advantage that the large mass of
cases with $\bar{D}_{JS}=1$ are spread out over a range of values of
$\bar{D}_{\rm effective}$ providing a little more information.  We
found minor improvements in clustering with the latter metric, and so
shall use this in what follows.

\begin{figure}[th]
    \centering 
    \includegraphics[width=\linewidth]{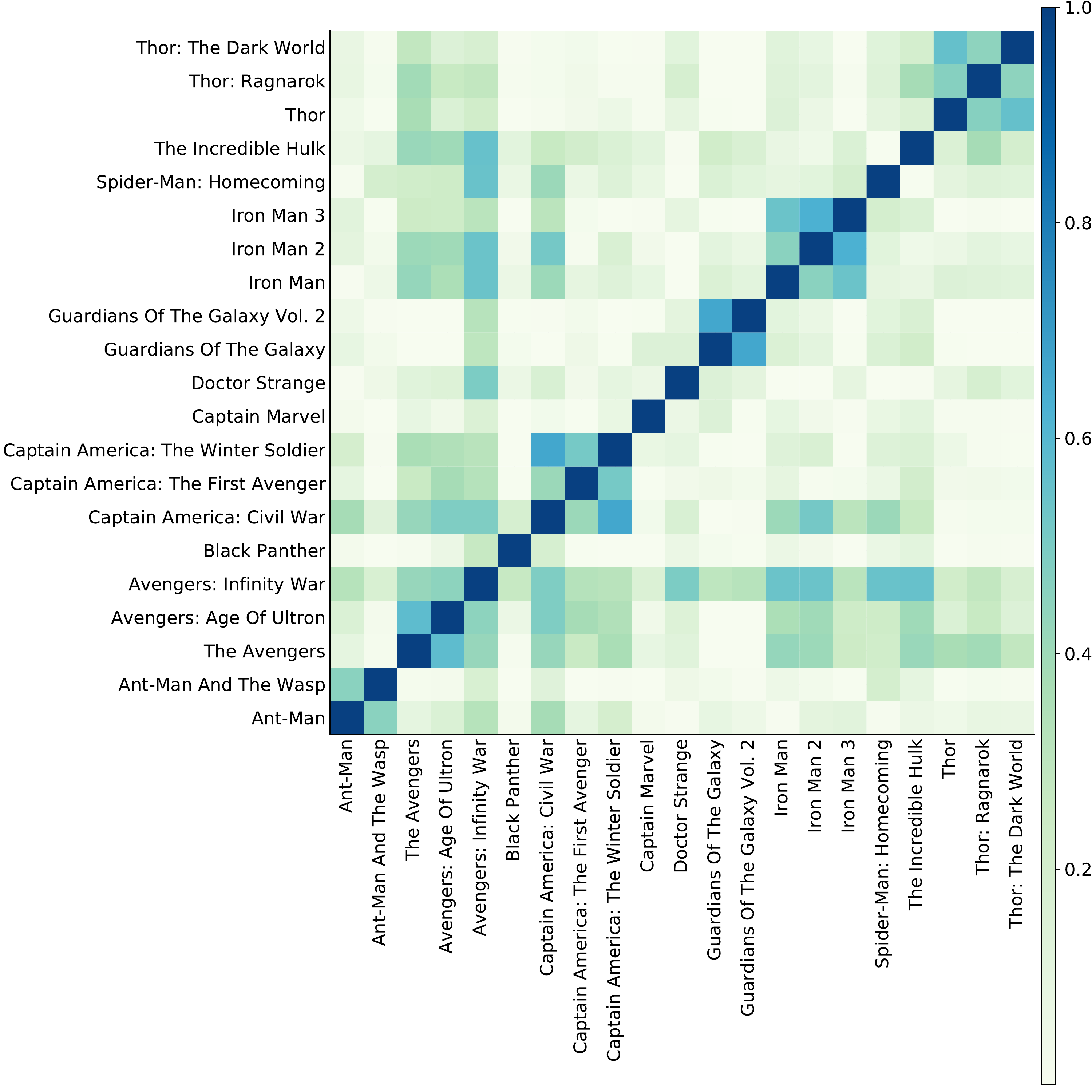}
    \caption{Heat map of normalised similarities $\bar{S}_{\rm effective}$
      between pairs of movies. Evident are blocks of movies
      corresponding to the major sub-sequences, \eg the Thor movies or
      the Iron Man movies. Also noticeable is the sharing of cast
      between the Avengers (team-up) movies and many of the others. }
    \label{fig:distance}
\end{figure}

In practical terms the metric performs exactly as you might
expect. \autoref{fig:distance} shows a heat-map of the similarities
$\bar{S}$. Obvious clusters emerge between directly related movies,
\eg the Thor sequence.

\begin{figure}[p]
    \centering 
    \includegraphics[width=\linewidth]{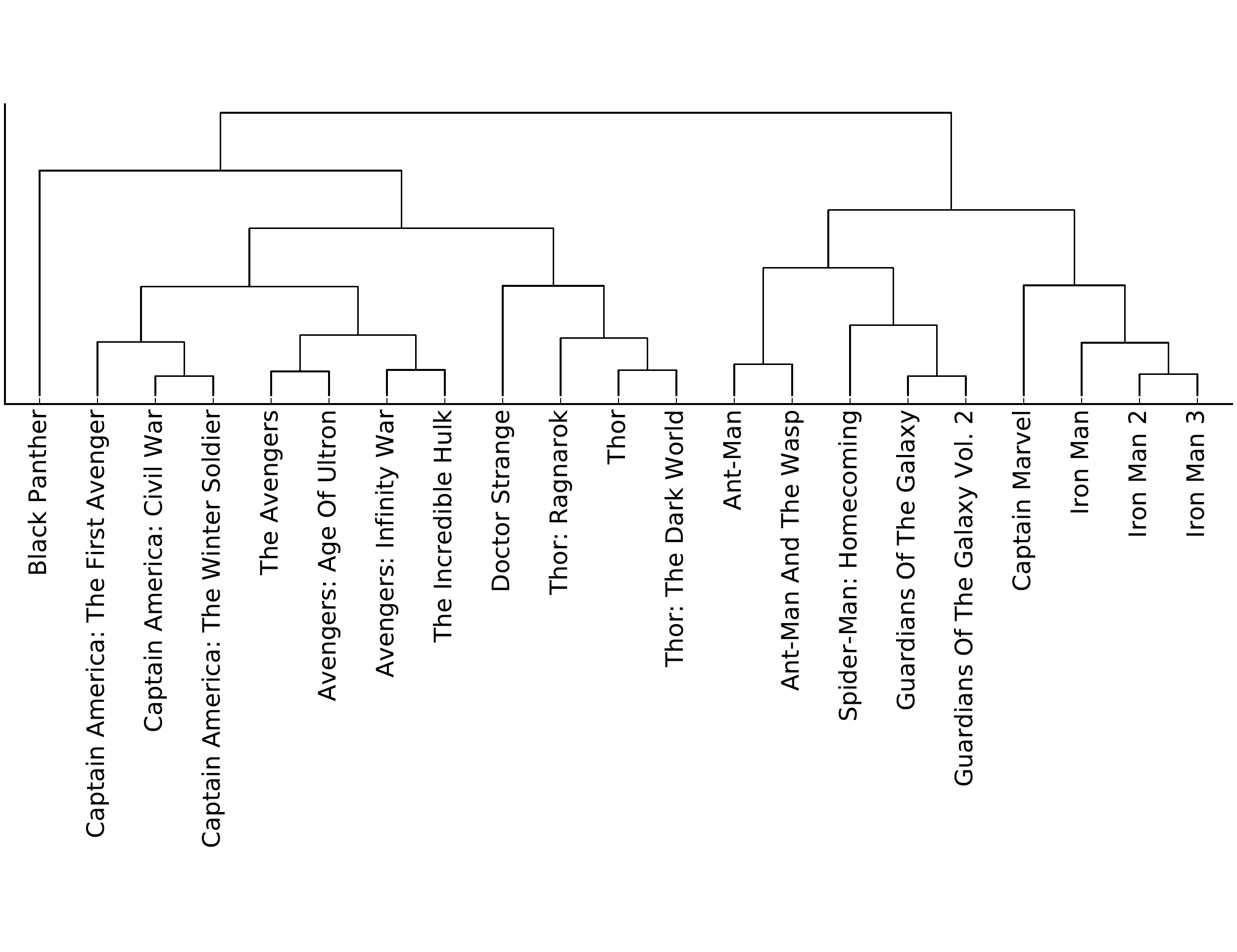}
    \vspace{-9mm}
    \caption{Dendrogram derived from hierarchical clustering of the
      movies based on the dis-similarities $D_{A,B}$}
    \label{fig:tree}
\end{figure}

We can draw out these clusters, for instance, using a hierarchical
clustering technique. The dendrogram for the method used is presented
in \autoref{fig:tree}. If we set the number of clusters to be 10
then we match the obvious groupings of the movies by sub-sequence.

\begin{figure}[p]
    \centering 
    \includegraphics[width=\linewidth]{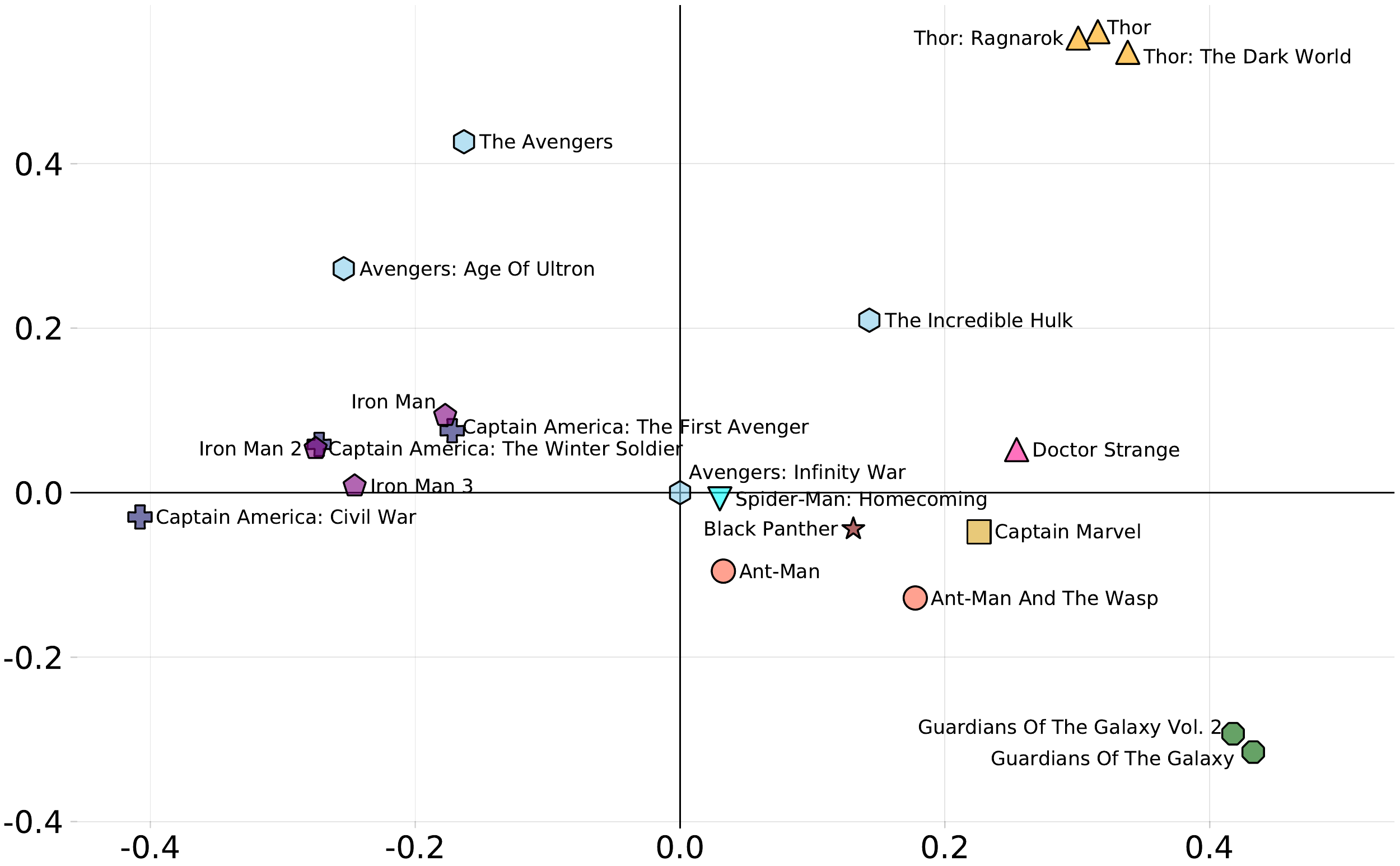}
    \vspace{2mm}
    \caption{MDS projection into a 2D space based on the cast
      dissimilarities. Note that a small translation has been applied
      to place \emph{Avengers: Infinity War}  at the origin. }
    \label{fig:mds}
\end{figure}

We illustrate this clustering by performing (classical) MDS on the
distance matrix $D$, to obtain projection of the movies into local
coordinates in a 2D Euclidean space. The coordinates are shown in
\autoref{fig:mds}, which also shows the clustering results. The
original metric space being embedded is highly non-Euclidean (many of
the distances are near 1), and so the 2D embedding is a poor
approximation in some ways, but it does allow visualisation of the
clusters. 

Of note: 
\begin{itemize}

\item the upper-right quadrant is the Thor/Hulk quadrant (remember that
  the Hulk plays a major role in \emph{Thor: Ragnarok}.

\item the upper-left quadrant is an Avengers/Iron Man/Captain America
  cluster. These movies are tightly wound together so it is no surprise
  that they appear together in this quadrant. 

\item The lower half holds a more miscellaneous collection of movies,
  though the two Guardians movies cluster tightly (again
  unsurprisingly). 

\end{itemize}
Note that although the Captain America movies are in a separate
cluster with the Iron Man/Avengers cluster, they appear to
overlap. This indicates the limitations of projecting into a 2D space,
the 3D projection more clearly separates these clusters, but is hard
to illustrate here. 

Informally, we can see in the plot a left-to-right division between
the more technology-derived superheros (Iron Man and Captain America)
vs those that come from a magical or alien background (Doctor Strange,
Thor and the Guardians). The placement of the Hulk and Ant-Man towards
the right is somewhat of a surprise from this point of view, but
perhaps reflects that sometimes superpowers are really magic explained
explained as technology.

It is tempting to speculate how other movies will embed in this
space. We might expect \emph{Avengers: End Game} to hold a similar or
even more central position than \emph{Avengers: Infinity War}.
\emph{Spider Man: Far From Home} is not released, but there is a
version of its cast list available, and it suggests it will sit close
to the existing Spider-Man movie.  We might speculate on hypothetical
sequels to \emph{Black Panther}, \emph{Captain Marvel}, or the
Guardians (likely close to the existing movies).

\section*{Conclusion and Future Work}

Measuring the number of characters in the a movie or franchise such as
the Marvel Cinematic Universe is an interesting and novel challenge.
While naively seen as easy, there is a need for a robust metric of
character count that incorporates character participation. Such a
metric is less sensitive to noise in the input data, and harder for a
commercial entity to ``game.''

The use of Shannon-entropy based measures built from frequency
distributions of participation provides a metric based on ``ecological
diversity,'' which is an intuitive, robust and useful metric.  The
robustness is shown through the comparisons of the metric under
constructions from dialogue and visual conflicts separately, and
alternative transcriptions of the same films.  The usefulness of this
metric is explored with respect to the success of the movies measured
via their profitability and shown to have a strong correlation.  A
generalisation of this metric using Jenson-Shannon divergence allows a
measure of similarity between movies that gives rise to a clean
clustering of movies, supporting the notion of usefulness of this
metric as a tool for the taxonomy of movies.

The usefulness of this metric does not end in the Marvel Universe and
possible extensions of this work are widespread. The presented metric
and its generalisation could be explored in relation to other
cinematic universes, or TV shows, where data is available. Further,
the metric has been applied to the Shannon-entropy of frequency
distributions, however, this metric could be applied to a variety of
other sources with relevance to character sentiments, transitions
between evil and good characters, or the entropy of the speaker
sequence as a Markovian process.

\section*{Supporting information}

\paragraph*{S1 Fig.} The numbers of credited characters in the MCU movies.
    \includegraphics[width=\linewidth]{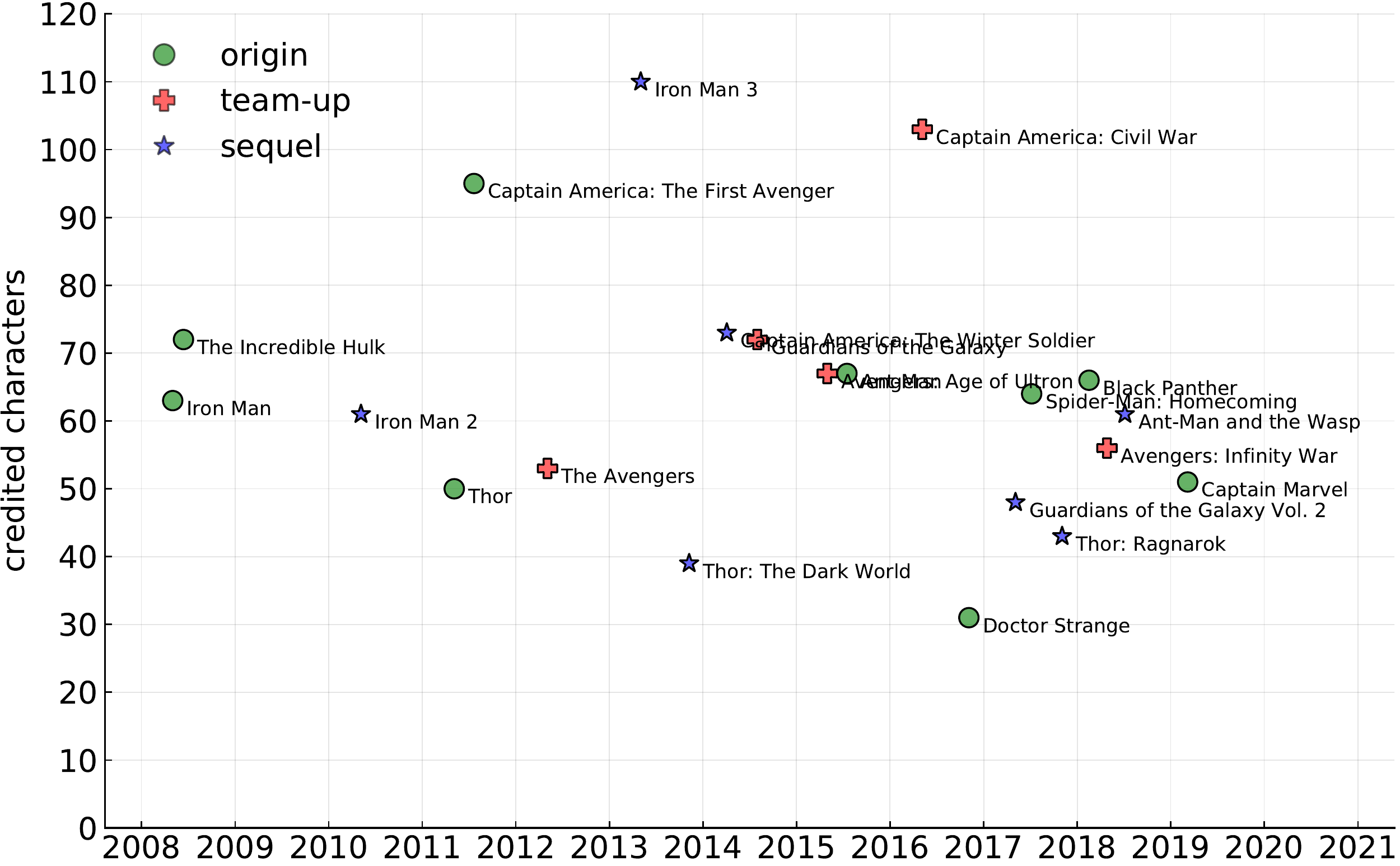}

\paragraph*{S2 Fig.} The numbers of named characters in the MCU movies.
    \includegraphics[width=\linewidth]{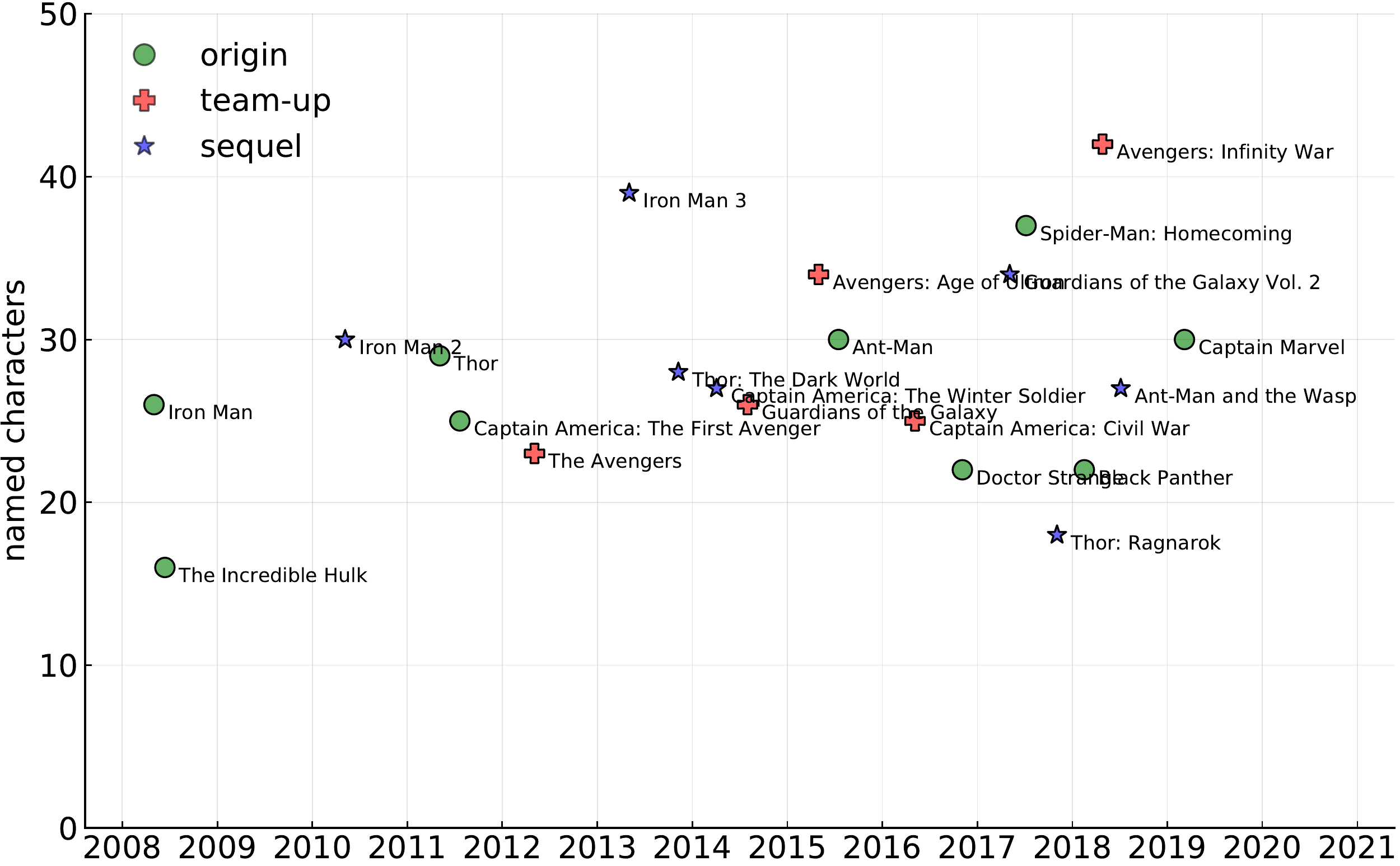}

\paragraph*{S3 Fig.} Jenson-Shannon similarities (compare to \autoref{fig:distance}).
    \includegraphics[width=\linewidth]{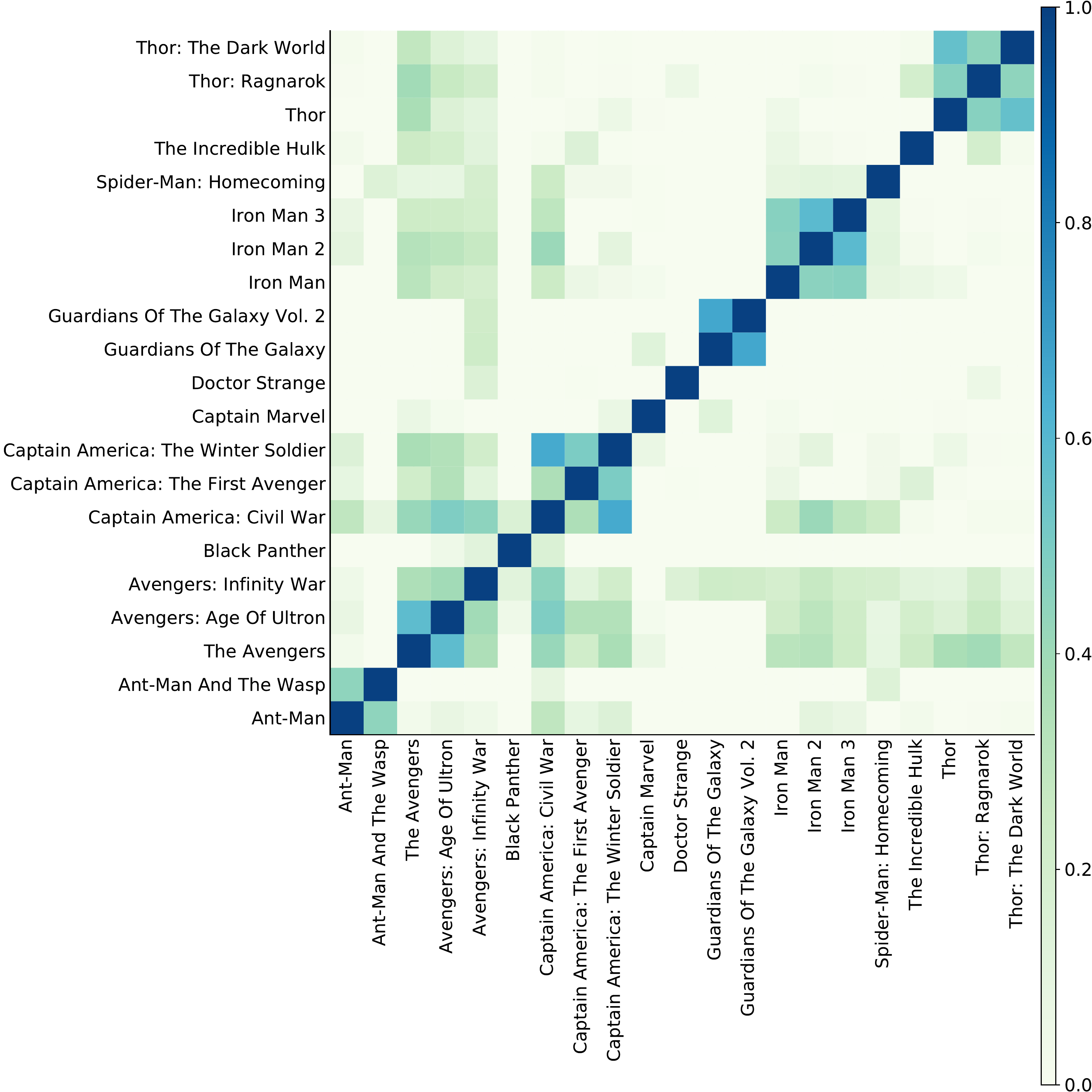}

\nolinenumbers

\bibliography{distances,entropy,marvel,dialogue,related}

\appendix

\end{document}